# Universal Dynamic Conductivity and Quantized Visible Opacity of Suspended Graphene

R.R. Nair[1], P. Blake[2], A.N. Grigorenko[1], K.S. Novoselov[1], T.J. Booth[2], T. Stauber[3], N.M.R. Peres[3] & A.K. Geim[1]

[1]*Manchester Centre for Mesoscience & Nanotechnology, University of Manchester, M13 9PL, Manchester, UK*
[2]*Graphene Industries Ltd, 32 Holden Avenue, M16 8TA, Manchester, UK*
[3]*Department of Physics, University of Minho, P-4710-057, Braga, Portugal*

We show that the optical transparency of suspended graphene is defined by the fine structure constant, $\alpha = e^2/\hbar c$, the parameter that describes coupling between light and relativistic electrons and is traditionally associated with quantum electrodynamics rather than condensed matter physics. Despite being only one atom thick, graphene is found to absorb a significant ($\pi\alpha = 2.3\%$) fraction of incident white light, which is a consequence of graphene's unique electronic structure. This value translates directly into universal dynamic conductivity $G = e^2/4\hbar$ within a few % accuracy.

## Introduction

There is a small group of phenomena in condensed matter physics, which are defined only by the fundamental constants and do not depend on material parameters. Examples are the resistivity quantum $h/e^2$ that appears in a variety of transport experiments, including the quantum Hall effect and universal conductance fluctuations, and the magnetic flux quantum $h/e$ playing an important role in the physics of superconductivity (here $h$ is the Planck constant and $e$ the elementary charge). By and large, it requires sophisticated facilities and special measurement conditions to observe any of these phenomena. Here, we show that such a simple observable as the visible transparency of graphene [1] is defined by the fine structure constant, $\alpha$. Our results are in agreement with the theory of ideal two-dimensional (2D) Dirac fermions [2,3] and its recent extension into visible optics [4], which takes into account the triangular warping and nonlinearity of graphene's electronic spectrum.

## Fabrication of graphene membranes

Large graphene crystals were prepared by micromechanical cleavage of natural graphite flakes (www.graphit.de) on top of an oxidized Si wafer [5] with an additional thin layer of PMMA [6]. The latter significantly improved adhesion and allowed us to make graphene monocrystals that could easily exceed 100 μm in size. We usually use NITTO tape to minimize contamination by adhesive residues. Single-, double- or few-layer crystallites were identified in an optical microscope due to their different contrast that increases with increasing the number of layers [6]. The number of layers was also verified with atomic force and Raman microscopy [7].

A perforated ≈20-μm-thick copper-gold film was then deposited on top of the found crystallites. The films normally had 9 small apertures with diameters 20, 30 and 50 μm as shown in the inset of Fig 1. The graphene crystallites were aligned against the apertures to cover them completely or partially. The Cu/Au film also served as a support structure (scaffold) and was 3 mm in diameter so that it could be used in standard holders for transmission electron microscopy (TEM). At the final stage of microfabrication, the scaffold was lifted off by dissolving the sacrificial PMMA layer, which left graphene attached to the scaffold (the use of a critical point dryer was essential).

The resulting devices could easily be handled and transferred between different measurement facilities. The developed technique allows a reliable and routine fabrication of practically macroscopic graphene membranes suitable for optical, electron-microscopy, micromechanical and other studies (our success rate in making final devices is >50%). So large one-atom-thick membranes were previously inaccessible and present a significant experimental advance with respect to the earlier fabrication procedures that largely relied on chance and allowed graphene membranes of only a few microns in size [8,9].

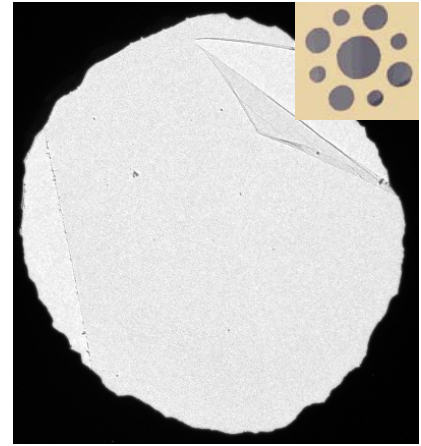

Figure 1. TEM micrograph of a 30 μm aperture covered by graphene. Because membranes that uniformly cover the whole aperture are essentially invisible, we show a sample with defects: there is a gap on the left and folded area (top right). The inset illustrates our design: a metal scaffold has several sub-mm apertures with graphene crystallites placed on top of them.

**Optical measurements**
To measure the optical spectra, we used a xenon lamp (wavelength $\lambda$ between 250 and 1200nm) as a light source and focused its beam on graphene membranes. The transmitted light intensity was measured by Ocean Optics HR2000 spectrometer. The recorded signal was then compared with the one obtained by directing the light beam through either an empty space or, as a double check, another aperture of the same size but without graphene. Typical experimental data are shown in Figure 2 by open circles. Here, to reduce the measurement noise below 0.1%, we have averaged the spectral curves over intervals $\Delta\lambda$ of 10 nm. The measurements yield graphene's opacity of 2.3% which is practically independent of wavelength $\lambda$. An interesting alternative method to measure opacity of graphene was to use membranes that only partially covered the apertures and take their images in an optical microscope (we used Nikon Eclipse LV100). In this case, opacities of different areas can be compared directly. The images taken by a high-quality grey-scale camera (Nikon DS2MBW) were then analyzed, and relative intensities of the light transmitted through different areas were calculated. Figure 3 shows an image of one of such samples in transmitted white light. The line scan across the image qualitatively illustrates changes in the observed light intensity. This method has also allowed us to measure graphene's transmittance as a function of $\lambda$. To this end, we used 22 different narrow-band-pass filters for transmission (back-side) illumination. Examples of such spectroscopy for graphene and its bilayer are shown in Fig. 4. Results of the two measurement techniques are compared in Figure 2 (circles versus squares) and show nearly the same accuracy. Note that the use of an optical microscope is possible for graphene membranes because they mostly absorb light with only a minute portion of it being reflected (<0.1%). The latter ensures that the opacity of graphene is practically independent of the numerical aperture and magnification (this was carefully checked experimentally and is in agreement with our calculations).

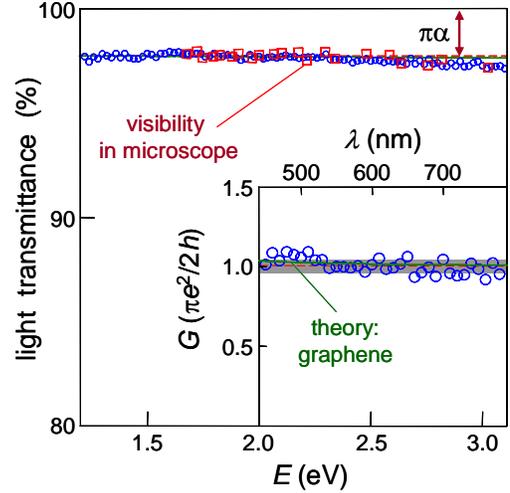

Figure 2. Transmittance spectrum of graphene over a range of photon energies $E$ from near-infrared to violet. The blue open circles show the data obtained using the standard spectroscopy for a membrane that completely covered a 30 μm aperture. For comparison, we show the spectrum measured using an optical microscope (squares). The red line indicates the opacity of $\pi\alpha$. Inset: Dynamic conductivity $G$ of graphene for visible wavelengths (symbols) recalculated from the measured $T$. The green curves in both main figure and inset show the expected theoretical dependences (see further), in which $G$ varies between 1.01 and 1.04 of $G_0 \equiv e^2/4\hbar$ for this frequency range. The red line and the gray area indicate the statistical average of our measurements and their standard error, respectively: $G/G_0 = 1.01 \pm 0.04$.

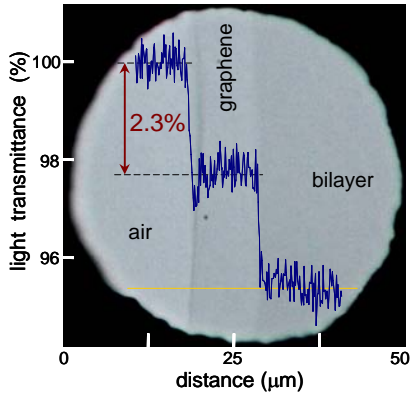

Figure 3. Looking through one-atom-thick crystals. Optical photograph of a 50 μm aperture partially covered by graphene and its bilayer. The line scan profile shows the intensity of transmitted white light along the yellow line.

Both approaches to measure graphene transmittance spectra show a deviation from a constant opacity for $\lambda < 500$ nm (photon energy $E > 2.5$eV), and the same is valid for bilayer graphene (see Figs. 2 and 4). Such rapid deviations are not expected in theory (see below). We have investigated this spectral feature further and found that its origin lies in surface contamination of graphene membranes by hydrocarbons. Graphene is extremely lipophilic and hydrocarbon contamination is practically impossible to avoid for samples exposed to air (a hydrocarbon layer partially covering graphene is always found in TEM; see, for example, [8]). To this end, we annealed our membranes in a hydrogen-argon atmosphere at 200C° [10], which significantly improved their cleanliness, as observed in TEM by using the membranes immediately after their annealing. The annealing is found to significantly weaken the downturn in the violet-light transmittance but did not affect the spectra for $\lambda$ >500nm, which indicates that hydrocarbons are indeed responsible for this additional opacity (or, at least, most of it). Here we note that many polymer (hydrocarbon) materials, especially those used in lithography, have an absorption edge in violet light. Alternatively, we speculate that it could be a tail of the plasmon resonance expected at $E \approx$5eV, which is broadened by surface contaminants. For thicker layers of a resist residue left on graphene, we observed a significant increase in opacity (by a factor of 2) over the whole frequency range but such contamination is accompanied by an increase in reflectivity that can be detected in an optical microscope. We have also measured the white-light opacity for multilayer graphene membranes and found it to increase with membranes' thickness so that each graphene layer adds another $\approx$2.3% as shown in see Figure 5.

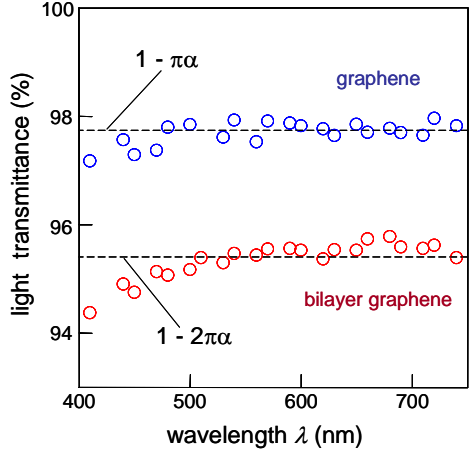

Figure 4. Transmittance spectra of single and bilayer regions of the sample shown in Fig. 3. The transmittance was measured by analyzing images taken in an optical microscope when the membrane was illuminated through narrow-band filters.

**Universal dynamic conductivity of graphene**
Optical properties of thin films are commonly described in terms of dynamic or optical conductivity $G$. For a 2D Dirac spectrum with a conical dispersion relation $\varepsilon = \hbar v_F |\mathbf{k}|$ ($v_F \approx 10^6$m/s is the Fermi velocity and $\mathbf{k}$ the wavevector), $G$ was theoretically predicted [2,3] to exhibit a universal value $G_0 \equiv e^2/4\hbar$, if the photon energy $E$ is much larger than both temperature and Fermi energy $\varepsilon_F$. Both conditions are stringently satisfied in our visible-optics experiments. The universal value of $G$ also implies that all optical properties of graphene (its transmittance $T$, absorption $P$ and reflection $R$) can be expressed through fundamental constants only ($T$, $P$ and $R$ are unequivocally related to $G$ in the 2D case). In particular, it was noted by Kuzmenko et al [11] that $T = (1+2\pi G_0/c)^{-2} = (1+0.5\pi\alpha)^{-2} \approx 1-\pi\alpha$ for the normal light incidence. We emphasize that – unlike $G$ – both $T$ and $R$ are observable quantities that can be measured directly by using graphene membranes.

To find accurate absolute values of $T$ and $G$, in the analysis shown in Fig. 2, we have omitted the part of the transmittance spectra at $\lambda$ <500 nm, which as discussed above was affected by hydrocarbon contamination. Also, our apparatus noise was somewhat higher in the infrared region so that, after including the infrared data, the statistical error over the whole spectrum usually grew rather than decreased. Accordingly, we restricted the analysis to $\lambda$ <800nm to maximize the accuracy. As a result, we have found $T \approx$97.7% with an accuracy of ±0.1%. The related analysis for $G$ yields $G \approx 1.01 G_0$ over the white-light region (450 nm <$\lambda$ <800 nm) and the standard error of ±4% (see Fig. 2).

The approximation of 2D Dirac fermions is valid for graphene only close to the Dirac point and, for higher energies $\varepsilon$, one has to take into account such effects in graphene's band structure as triangular warping and nonlinearity [4]. The triangular warping is significant even for $E$<1 eV, and there is little left of the linear Dirac spectrum at $\varepsilon$ approaching 5 eV. Therefore, for visible energies of 2 to 3 eV, which are already comparable with the nearest-neighbor hopping energy $t \approx 3$ eV, one may expect the breakdown of the Dirac-fermion approximation used in the calculations of $G_0$. Accordingly, the only earlier theoretical analysis [11] that did take into account the finite-$E$ corrections was still limited to the infrared region. For the purpose of our experiments, we have extended the theory to visible frequencies and, also, took into account the next-near-neighbor hopping. The latter was found to result only in minute corrections (see ref. [4] for details). Figure 6 shows the calculated dynamic conductivity $G$ and light transmittance $T$ with the finite-$E$ effects included. One can see a noticeable increase in $G$ at a finite $E$ with respect to its idealized value of $e^2/4\hbar$ but the corrections still do not exceed 2% for green light. Note that, in the infrared region, the corrections do not disappear but decrease relatively slowly (as $\propto E^2$), until one needs to take into account finite temperature and $\varepsilon_F$ [2-4]. Our calculations are also in quantitative agreement with the earlier results for $E \leq 1$ eV [11].

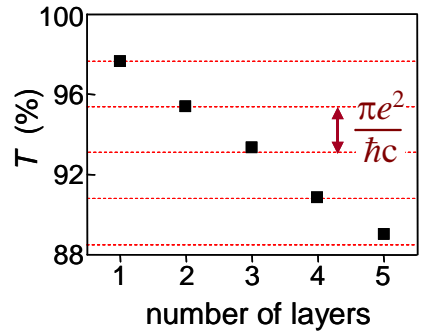

Figure 5. Transmittance of white light as a function of the number of graphene layers (squares). The dashed lines correspond to the reduction in intensity by $\pi\alpha$ with each added layer.

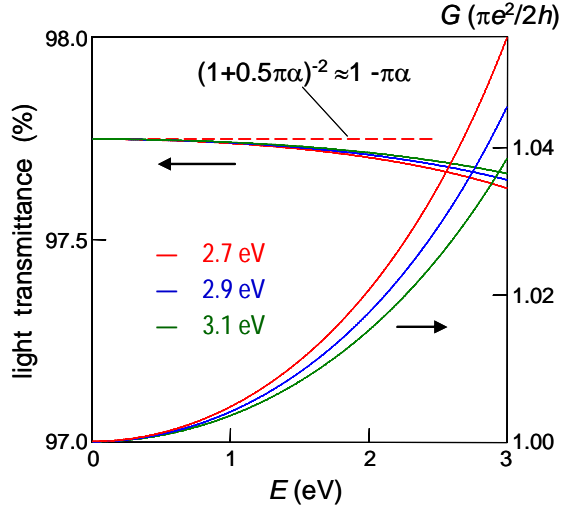

Figure 6. Dynamic conductivity as a function of photon energy $E$ for graphene, taking into account its triangular warping and nonlinearity at finite energies. The curves are given for 3 values of $t$ which cover the possible range expected for this hopping parameter. The corresponding curves for light transmittance are also shown. The red dashed line indicates the value for the idealized case of 2D Dirac fermions.

Now we turn our attention to few-layer graphene. It is surprising that its opacity is proportional to the number $N$ of layers involved, at least, to a good approximation for $N \leq 4$ (see Fig. 5). Indeed, electronic structures of the multilayer materials are different for different $N$. Generally, several energy bands are present for $N \geq 2$, and the interband distance is given by the energy of inter-plane hopping, $t_\perp \approx 0.3$ eV. This leads to complicated optical spectra with marked absorption peaks corresponding to interband transitions [11,12]. However, for visible photon energies $E \gg t_\perp$ the spectra significantly simplify so that up to corrections of the order of $(t_\perp/E)^2 \ll 1$ multilayer graphene can be considered as a stack of independent graphene planes [4]. This leads to the opacity $(1-T) \approx N\pi\alpha$, which was explicitly derived for bilayer graphene $N=2$ [4] and, also, is apparent from Figure 1 of ref. [12]. Further theoretical analysis is required for multilayer graphene.

**Absorption of light by 2D Dirac fermions**

Finally, we show how the universal value of graphene's opacity can be understood qualitatively, without calculating its dynamic conductivity. Let a light wave with electric field $\vec{\Theta}$ and frequency $\omega$ fall perpendicular to a graphene sheet of a unit area. The incident energy flux is given by $W_i = \frac{c}{4\pi}|\Theta|^2$. Taking into account the momentum conservation $k$ for the initial $|i\rangle$ and final $|f\rangle$ states, only the excitation processes pictured in Fig. 7 contribute to the light absorption. The absorbed energy $W_a = \eta\hbar\omega$ is given by the number $\eta$ of such absorption events per unit time and can be calculated by using Fermi's golden rule as $\eta = (2\pi/\hbar)|M|^2 D$ where $M$ is the matrix element for the interaction between light and Dirac fermions, and $D$ is the density of states at $\varepsilon = E/2 = \hbar\omega/2$ (see Fig. 7). For 2D Dirac fermions, $D(\hbar\omega/2) = \hbar\omega/\pi\hbar^2 v_F^2$ and is a linear function of $\varepsilon$.

The interaction between light and Dirac fermions is generally described by the Hamiltonian

$$\hat{H} = v_F \vec{\sigma} \cdot \vec{p} = v_F \vec{\sigma} \cdot (\hat{p} - \frac{e}{c}\vec{A}) = \hat{H}_0 + \hat{H}_{int}$$

where the first term is the standard Hamiltonian for 2D Dirac quasiparticles in graphene [1] and $\hat{H}_{int} = -v_F \vec{\sigma} \cdot \frac{e}{c}\vec{A} = v_F \vec{\sigma} \cdot \frac{e}{i\omega}\vec{\Theta}$ describes their interaction with electromagnetic field. Here $\vec{A} = \frac{ic}{\omega}\vec{\Theta}$ is the vector potential and $\vec{\sigma}$ the standard Pauli matrices. Averaging over all initial and final states and taking into account the valley degeneracy, our calculations yield

$$|M|^2 = |\langle f| v_F \vec{\sigma} \cdot \frac{e}{i\omega} \vec{\Theta} |i\rangle|^2 = \frac{1}{8} e^2 v_F^2 \frac{|\Theta|^2}{\omega^2}.$$

This results in $W_a = (e^2/4\hbar)|\Theta|^2$ and, consequently, absorption $P = W_a/W_i = \pi e^2/\hbar c = \pi\alpha$, both of which are independent of the material parameter $v_F$ that cancels out in the calculations of $W_a$. Also note that the dynamic conductivity $G \equiv W_a/|\Theta|^2$ is equal to $e^2/4\hbar$. Because graphene practically does not reflect light ($R \ll 1$ as discussed above), its opacity $(1 - T)$ is dominated by the derived expression for $P$.

In the case of a zero-gap semiconductor with a parabolic spectrum (e.g., bilayer graphene at low $\varepsilon$), the same analysis based on Fermi's golden rule yields $P = 2\pi\alpha$. This shows that the fact that the optical properties of graphene are defined by the fundamental constants is related to its 2D nature and zero energy gap and does not directly involve the chiral properties of Dirac fermions.

On a more general note, graphene's Hamiltonian $\hat{H}$ has the same structure as for relativistic electrons (except for coefficient $v_F$ instead of the speed of light $c$). The interaction of light with relativistic particles is described by a coupling constant, a.k.a. the fine structure constant. The Fermi velocity is only a prefactor for both Hamiltonians $\hat{H}_0$ and $\hat{H}_{int}$ and, accordingly, one can expect that the coefficient may not change the strength of the interaction, as indeed our calculations show.

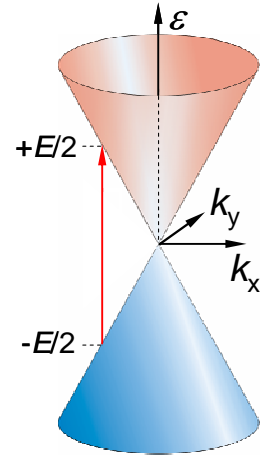

Fig. 7. Excitation processes responsible for absorption of light in graphene. Electrons from the valence band (blue) are excited into empty states in the conduction band (red) with conserving their momentum and gaining the energy $E = \hbar\omega$.

**Conclusions**

We have found that the visible opacity of suspended graphene is given by $\pi\alpha$ within a few percent accuracy and increases proportionally to the number of layers $N$ for few-layer graphene. Its dynamic conductivity at visible frequencies is remarkably close to the universal value of $(e^2/4\hbar)N$. The agreement between the experiment and theory is particularly striking because it was believed the universality could hold only for low energies (< 1eV) beyond which the electronic spectrum of graphene becomes strongly warped and nonlinear and the approximation of Dirac fermions breaks down. Our calculations show that finite-$E$ corrections are small even for visible light. Because of these corrections, a metrological accuracy for $\alpha$ would be difficult to achieve but it is remarkable that the fine structure constant can so directly and accurately be assessed practically by a naked eye.

*Acknowledgments* – We are grateful to Alexey Kuzmenko, Antonio Castro Neto, Philip Kim and Laurence Eaves for illuminating discussions.